\documentclass[aps,prc,12pt,superscriptaddress,showpacs,showkeys]{revtex4} 
\usepackage{epsfig,amsmath}
\newcommand{\eq}[1]{\begin{align} #1 \end{align}}

\usepackage{color}

\begin{document}


\title{ Jets propagation through a hadron-string medium\\
}

\author{V.P.~Konchakovski}
\affiliation{Institut f\"ur Theoretische Physik, Universit\"at Giessen, Germany}
\affiliation{Bogolyubov Institute for Theoretical Physics, Kiev, Ukraine}
\author{E.L.~Bratkovskaya}
\affiliation{Institut f\"ur Theoretische Physik, Universit\"at Frankfurt, Germany}
\affiliation{Frankfurt Institute for Advanced Studies, Frankfurt, Germany}
\author{W.~Cassing}
\affiliation{Institut f\"ur Theoretische Physik, Universit\"at Giessen, Germany}
\author{M.I.~Gorenstein}
\affiliation{Bogolyubov Institute for Theoretical Physics, Kiev, Ukraine}
\affiliation{Frankfurt Institute for Advanced Studies, Frankfurt, Germany}

\begin{abstract}
Di-jet correlations in nucleus-nucleus collisions are studied within
the Hadron-String-Dynamics (HSD) transport approach taking into account
the reaction of the medium on the jet energy loss nonperturbatively.  A
comparison with the STAR and PHENIX data in central Au+Au collisions at
the RHIC energy $\sqrt{s}=200$ GeV is performed differentially, i.e.
with respect to correlations in azimuthal angle $\Delta \phi$ and
pseudorapidity $\Delta \eta$. The HSD results do not show enough
suppression for the `away-side' jets in accordance with earlier
perturbative studies.  Furthermore, the `Mach-cone' structure for the
angle distribution in the `away-side' jet as well as `ridge' long
rapidity correlations in the `near-side' jet -- observed by the STAR
and PHOBOS Collaborations -- are not seen in the HSD results, thus
suggesting a partonic origin.
\end{abstract}

\pacs{24.10.Lx, 25.75.-q, 25.75.Bh, 25.75.Gz}

\keywords{heavy-ion collision, jets, ridge}

\maketitle




High transverse momentum ($p_T$) partons are  informative probes of the
high-energy-density matter created in relativistic nucleus-nucleus
(A+A) collisions. These partons loose a large fraction of their energy
during the early stage of A+A collisions before hadron formation. Such
an energy loss is predicted to lead to a phenomenon known as jet
quenching \cite{Baier:1996sk,Gyulassy:2003mc,Kovner:2003zj}.  The
underlying jet structure for particle production at high-$p_T$ can be
probed using  di-hadron correlations, which measure the associated
particle distribution in azimuthal angle $\Delta \phi$ and
pseudorapidity $\Delta\eta$ with respect to the high-$p_T$ `trigger'
particle.  The data on two-particle spectra in the high-$p_T$ region in
Au+Au collisions for the c.m.s. energy of the nucleon pair
$\sqrt{s}_{NN}=200$~GeV can be summarized as follows:  1) strong
suppression of the away-side hadrons (jet
quenching)~\cite{Adler:2002tq,phenix}; 2) specific `Mach-cone'
structure in azimuthal angle $\Delta \phi$ in the region of the
away-side jet ~\cite{phenix}; 3) long-range pseudorapidity $\Delta
\phi$ correlation (`ridge') in the region of the near-side jet
\cite{STAR,PHOBOS}.

These experimental observations have generated great interest. In
particular, various theoretical models are proposed to explain the
ridge phenomenon, including (a) longitudinal flow push
\cite{model_a}, (b) broadening of quenched jets in turbulent color
fields \cite{model_b}, (c) recombination between thermal and
shower partons \cite{model_c}, (d) elastic collisions between hard
and medium partons (momentum kick) \cite{model_d}, and (e)
particle excess due to QCD bremsstrahlung or color-flux-tube
fluctuations focused by transverse radial flow \cite{model_e}.
Furthermore, long-range correlations in $\Delta \eta$ might be
a consequence of string-like correlation phenomena.

In order to explore especially the latter conjecture of string-like
correlations, we use the microscopic HSD transport model
\cite{Ehehalt,Geiss,Cass99} for the study of di-jet correlations, which
employs dominantly early string formation in elementary reactions and
their subsequent decay.  We recall, that the HSD model allows to
explore  systematically the change in the dynamics from elementary
nucleon-nucleon to central nucleus-nucleus collisions in a unique way
without changing or introducing new model parameters.  Inelastic
hadron--hadron collisions with energies above $\sqrt s\simeq 2.6~GeV$
are described by the Fritiof model \cite{LUND} (including Pythia v5.5
with Jetset v7.3 for the production and fragmentation of jets
\cite{PYTHIA0}) whereas low energy hadron-hadron collisions are
modelled in line with experimental cross sections.  We stress that no
explicit parton cascading is involved in our present transport
calculations.

The previous HSD analysis of high-$p_T$ spectra in Ref.~\cite{jetHSD}
includes all model details and discusses the nuclear modification factor
$R_{AA}(p_T)$ as the function of $p_T$ and centrality.  In the earlier
studies jets were considered perturbatively and no back-coupling to the
medium has been incorporated.  In extension of the previous
investigations we now include the full evolution of jets in the
transport approach including the response of the medium, which is
important as the `Mach-cone' and `ridge' structures  are attributed to
medium evolution effects due to jet-medium interactions.  In the
current HSD calculations we use approximately 30$\times 10^6$ of p+p
inelastic collision events and 0.5$\times 10^6$ of central Au+Au
collisions with impact parameter $b=0$.

The di-jet correlations are measured as a function of azimuthal angle
$\Delta \phi$ and pseudorapidity $\Delta \eta$ between the trigger
and associated particles:
\eq{\label{distr3D}
C(\Delta\eta,\Delta\phi)=\frac{1}{N_{trig}}
  \frac{d^2N_{assoc}}{d\Delta\eta\ d\Delta\phi}~,
}
where $N_{trig}$ is the number of trigger particles. To obtain the
di-jet correlations one has to subtract a background distribution. In
our calculations we use the mixed events method which allows to
properly subtract the background by taking associated particles for
each trigger particle from another randomly chosen event.

\begin{figure}[t!]
\epsfig{file=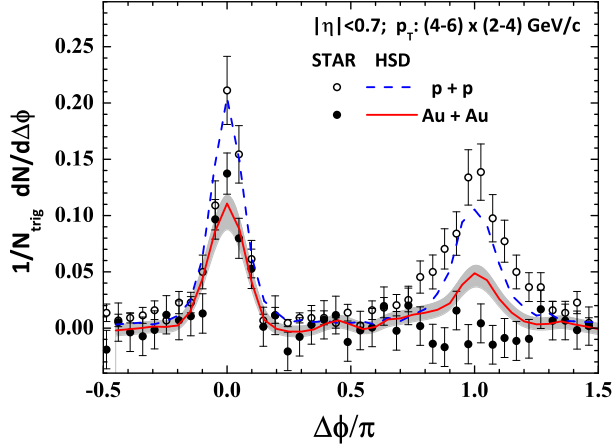,width=0.5\textwidth}
\caption{
Angular correlations of associated particles ($2<p_T^{assoc}<4$~GeV/c)
with respect to a trigger particle with
$p_T^{trig}>4$~GeV/c in p+p and in central Au+Au collision events within the HSD
transport approach in comparison to the STAR data~\cite{Adler:2002tq}.
The grey area corresponds to the statistical uncertainties of the HSD
calculations.}
\label{AngCorr0}
\end{figure}

\begin{figure}[t!]
\epsfig{file=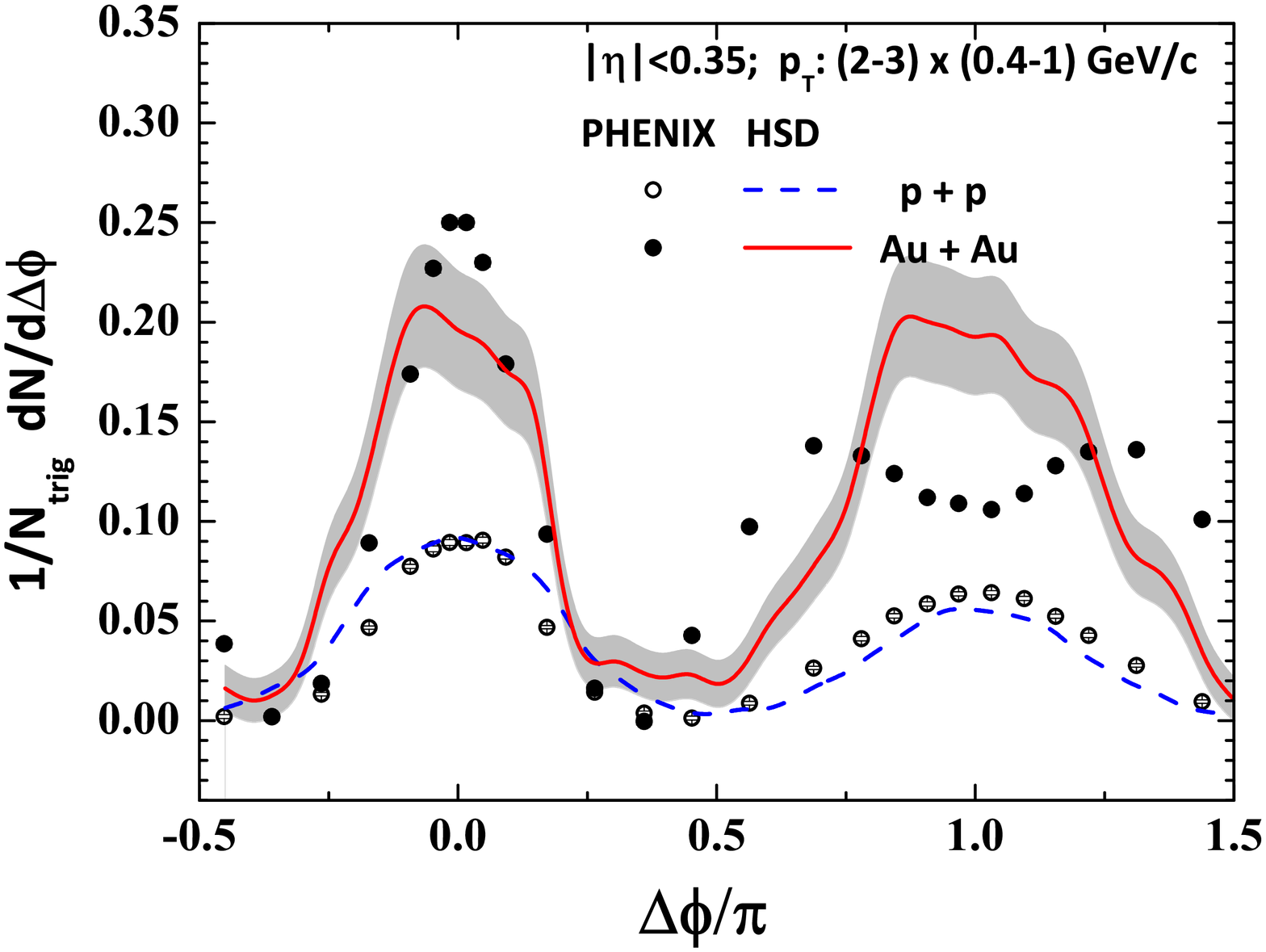,width=0.5\textwidth}
\epsfig{file=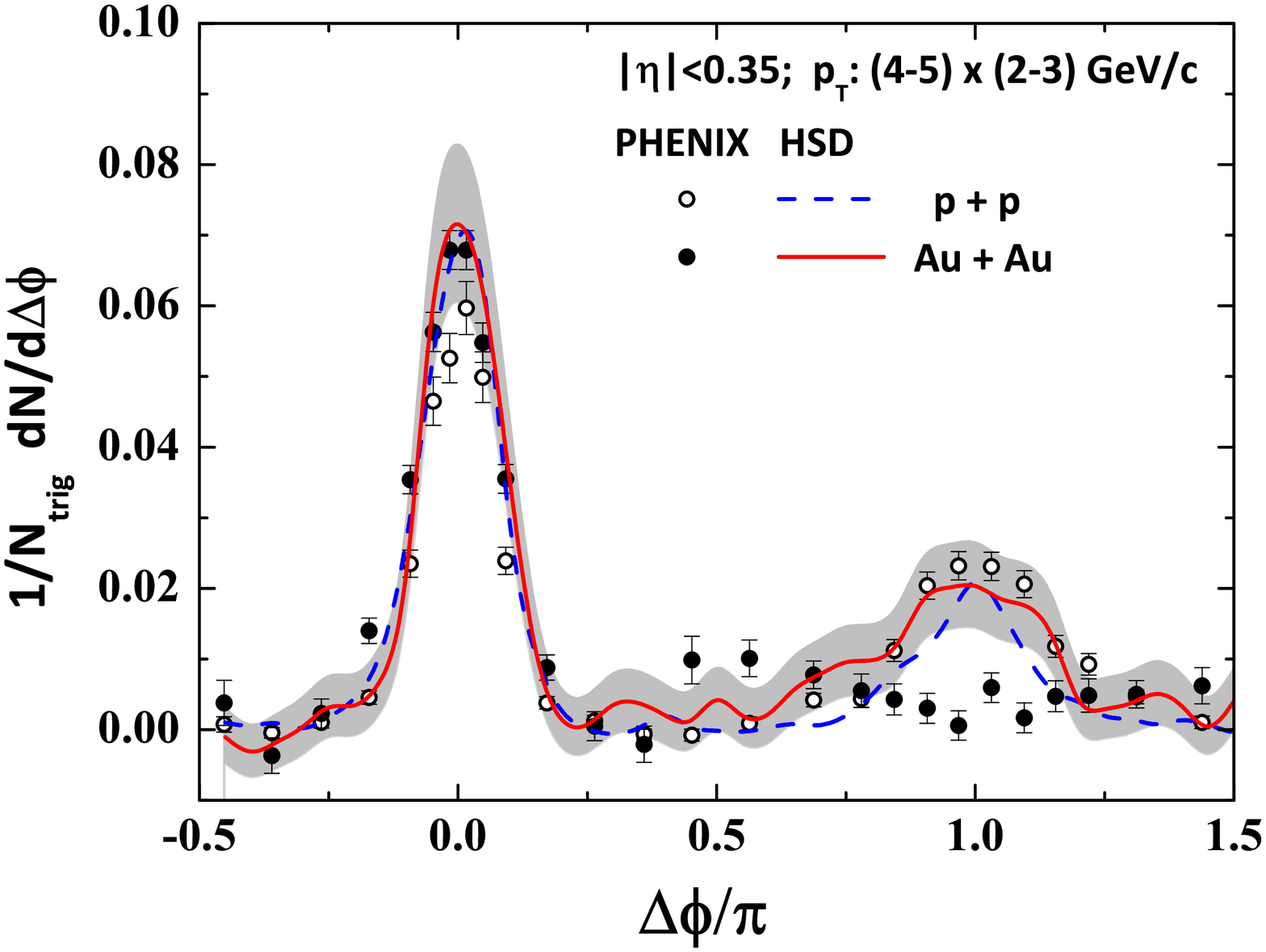,width=0.5\textwidth}
\epsfig{file=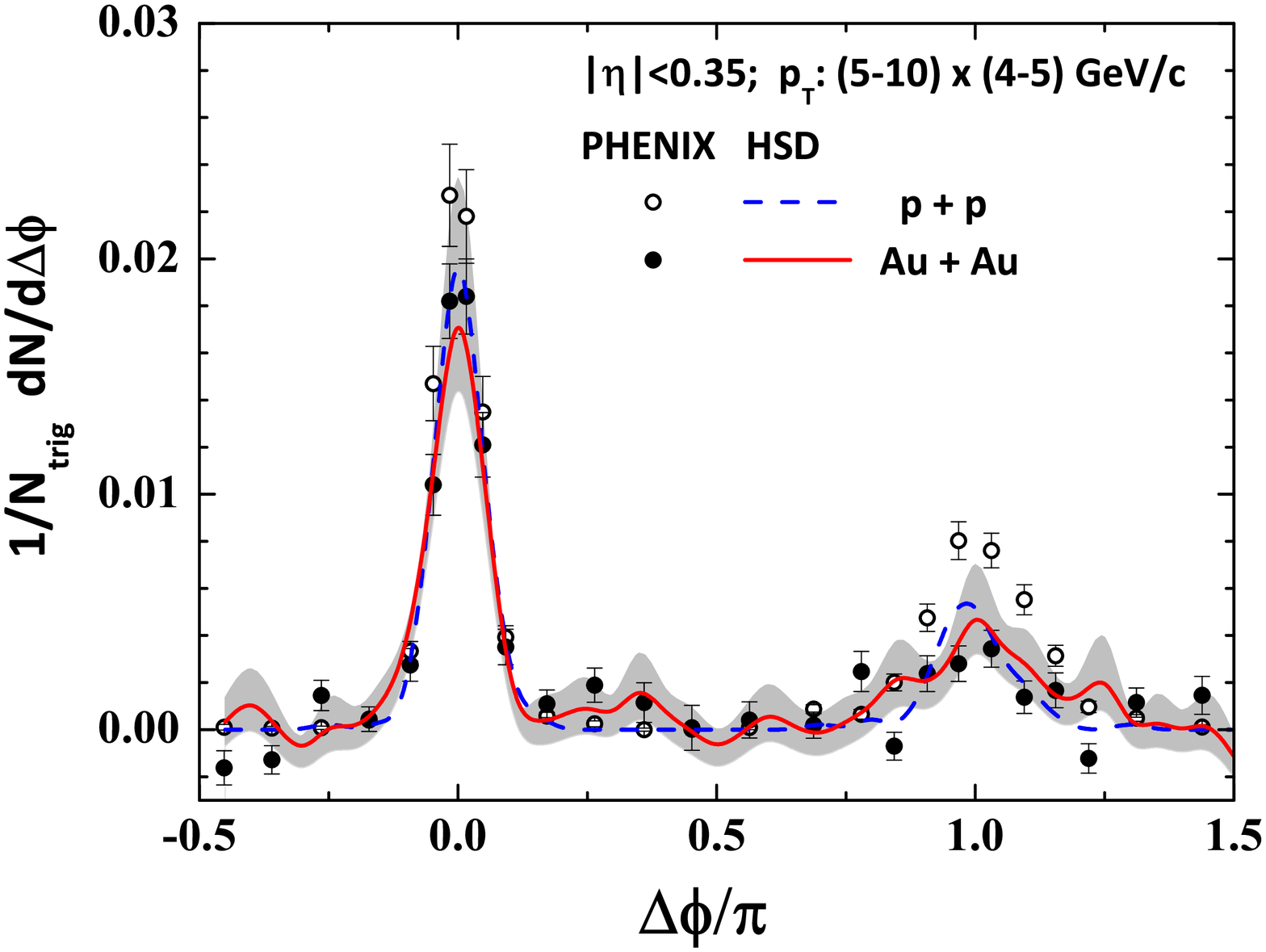,width=0.5\textwidth}
\caption{
Angular correlations of associated particles with different cuts for
$p_T^{trig}$ and $p_T^{assoc}$ in p+p and in central Au+Au collisions
within the HSD transport approach in comparison to the PHENIX
data~\cite{phenix}.  The grey areas correspond to the statistical
uncertainties of the HSD calculations.}
\label{AngCorr}
\end{figure}

Angular correlations of associated particles in p+p and central Au+Au
collisions with different cuts for $p_T^{trig}$ and $p_T^{assoc}$ are
shown in Figs.~\ref{AngCorr0} and \ref{AngCorr}.  There are two maxima:
the `near-side' and `away-side' peaks at $\Delta\phi=0$ and
$\Delta\phi=\pi$, correspondingly.  Fig.~\ref{AngCorr0} provides a
comaparison to the data from the STAR Collaboration with the cuts for
$p_T^{trig}$ and $p_T^{assoc}$ similar to those in the previous HSD
calculations \cite{jetHSD}. We find a good agreement between the
earlier perturbative~\cite{jetHSD} and current non-perturbative HSD
results.  Thus, one may conclude that a medium modification in this
kinematic region of $p_T^{trig}$ and $p_T^{assoc}$ is small in HSD. We
mention that the HSD results reasonably reproduce the data for p+p
collisions (within error bars).  For Au+Au central collisions HSD shows
clearly an insufficient suppression of the `away-side' peak at
$\Delta\Phi/\pi =1$. Note also that for the most central collisions the
experimentally observed suppression of single-particle spectra at
high-$p_T$ can not fully be described by HSD:
$R_{AA}^{HSD}(p_T)=0.35\div 0.4$ whereas $R_{AA}^{exp}(p_T)=0.2\div
0.25$ at $p_T>4$~GeV/c. Our first conclusion is that the hadron-string
medium is too transparant for high-$p_T$ particles (as already pionted
out in \cite{jetHSD}).

Fig.~\ref{AngCorr} corresponds to the data of the PHENIX Collaboration
with different cuts for $p_T^{trig}$ and $p_T^{assoc}$.  For the {\it
top} panel these cuts  are $p_T^{trig}=2\div 3$~GeV/c and
$p_T^{assoc}=0.4\div 1$~GeV/c.  This is the kinematic region where one
expects a strong medium response to the jet energy loss. The
experimental data show the presence of a `Mach-cone' structure in
azimuthal angle $\Delta \phi$ for the `away-side' jet. This structure
does not appear in the HSD simulations.

\begin{figure}[t!]
\epsfig{file=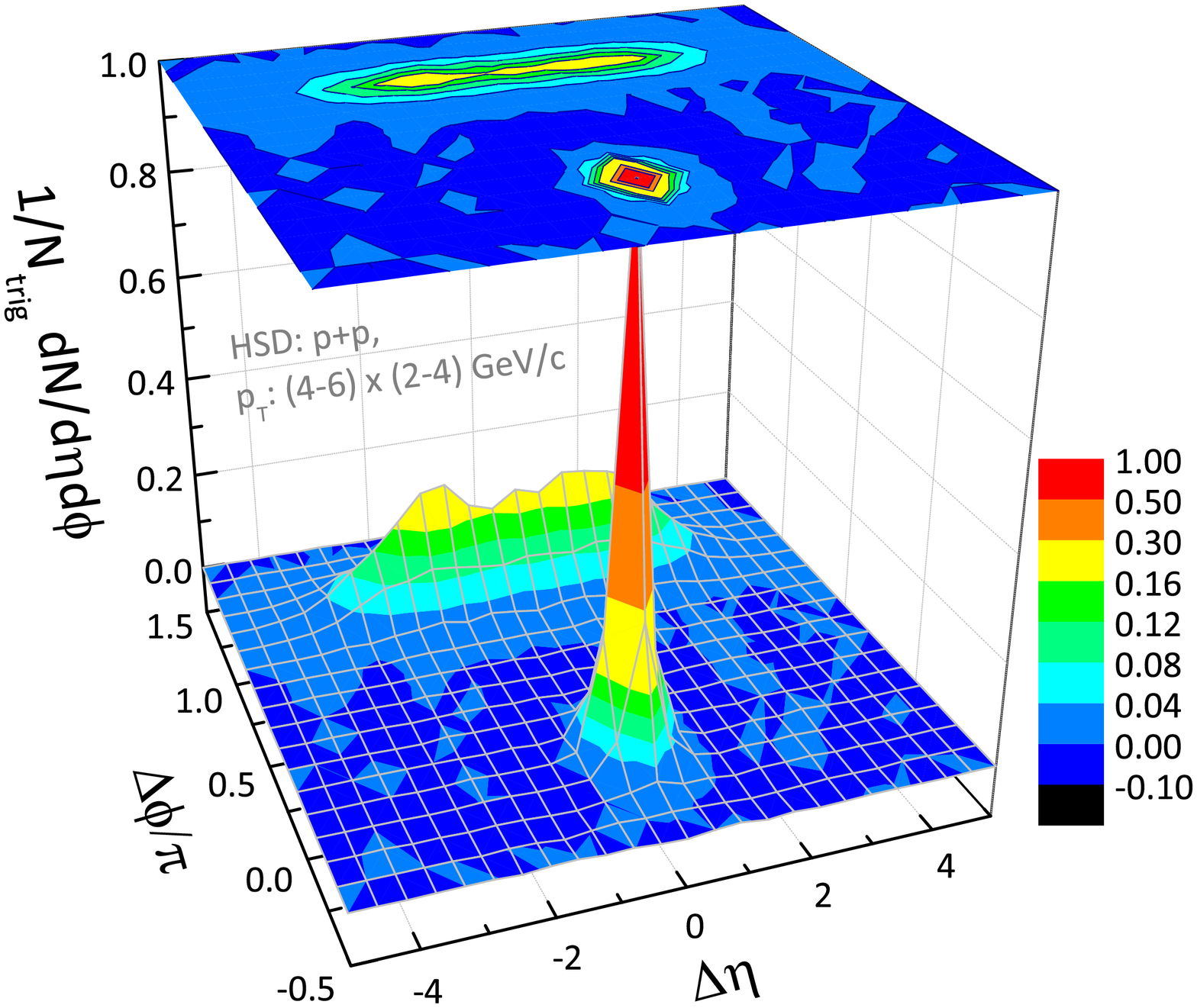,width=0.49\textwidth}
\epsfig{file=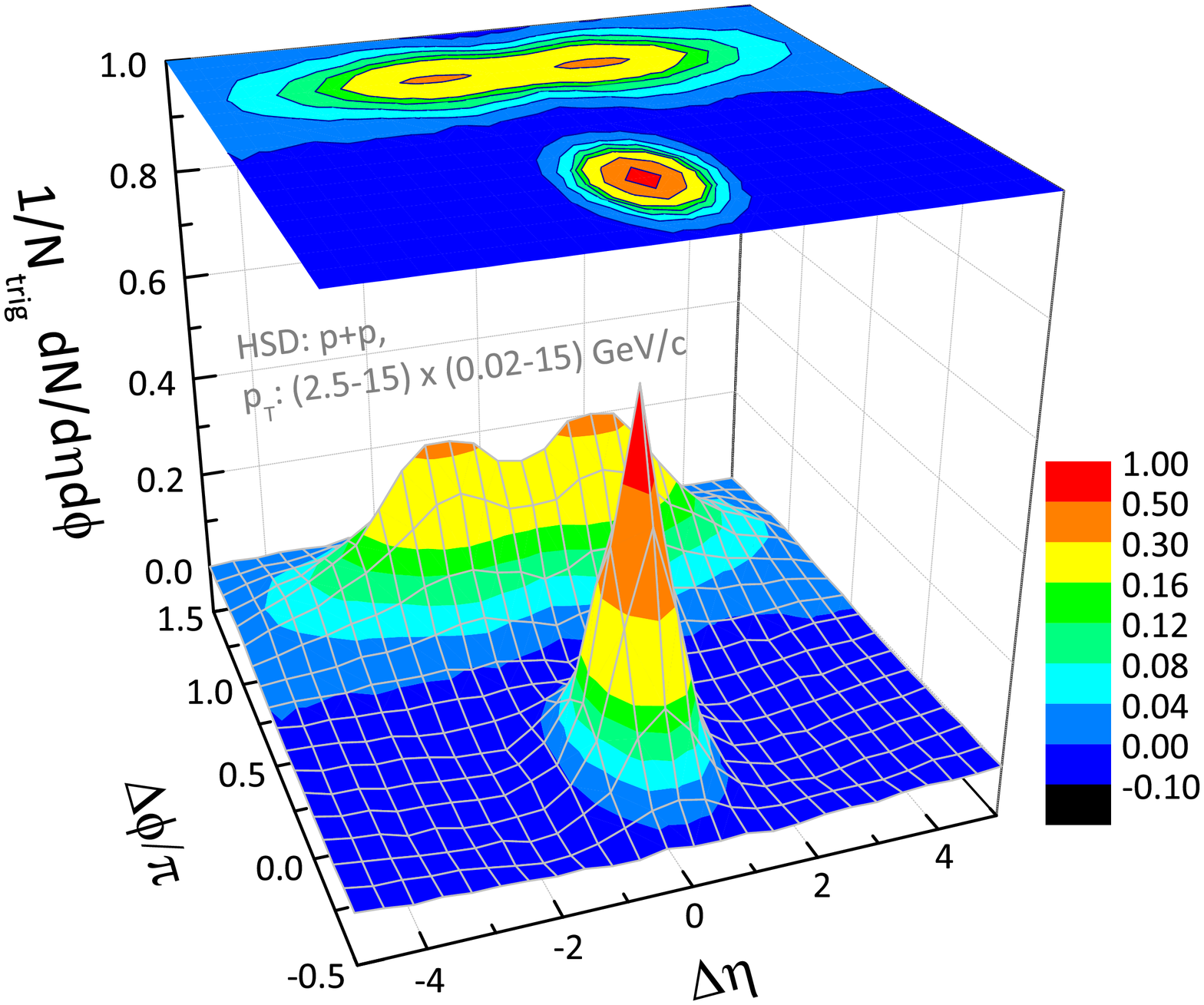,width=0.49\textwidth}
\epsfig{file=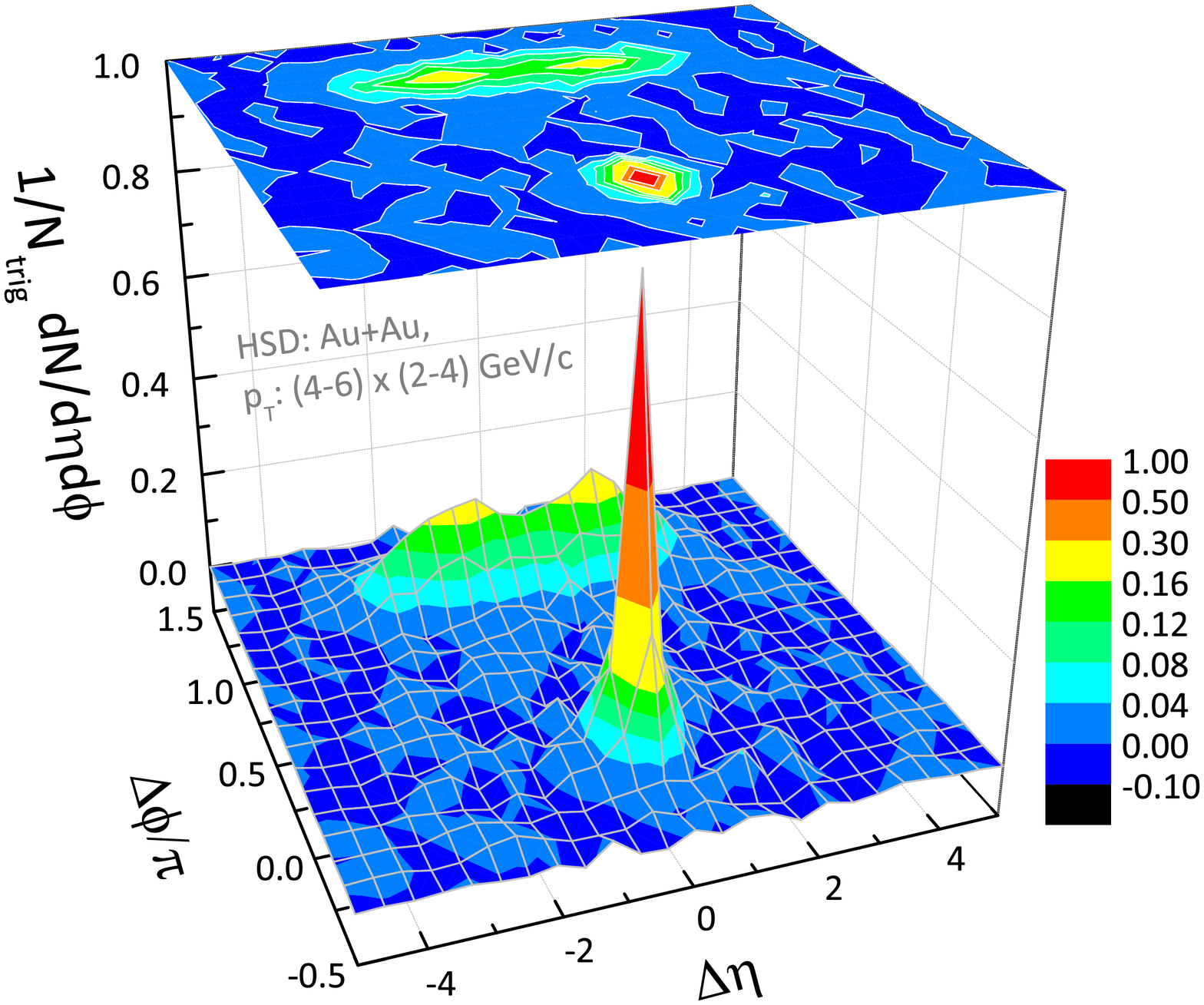,width=0.49\textwidth}
\epsfig{file=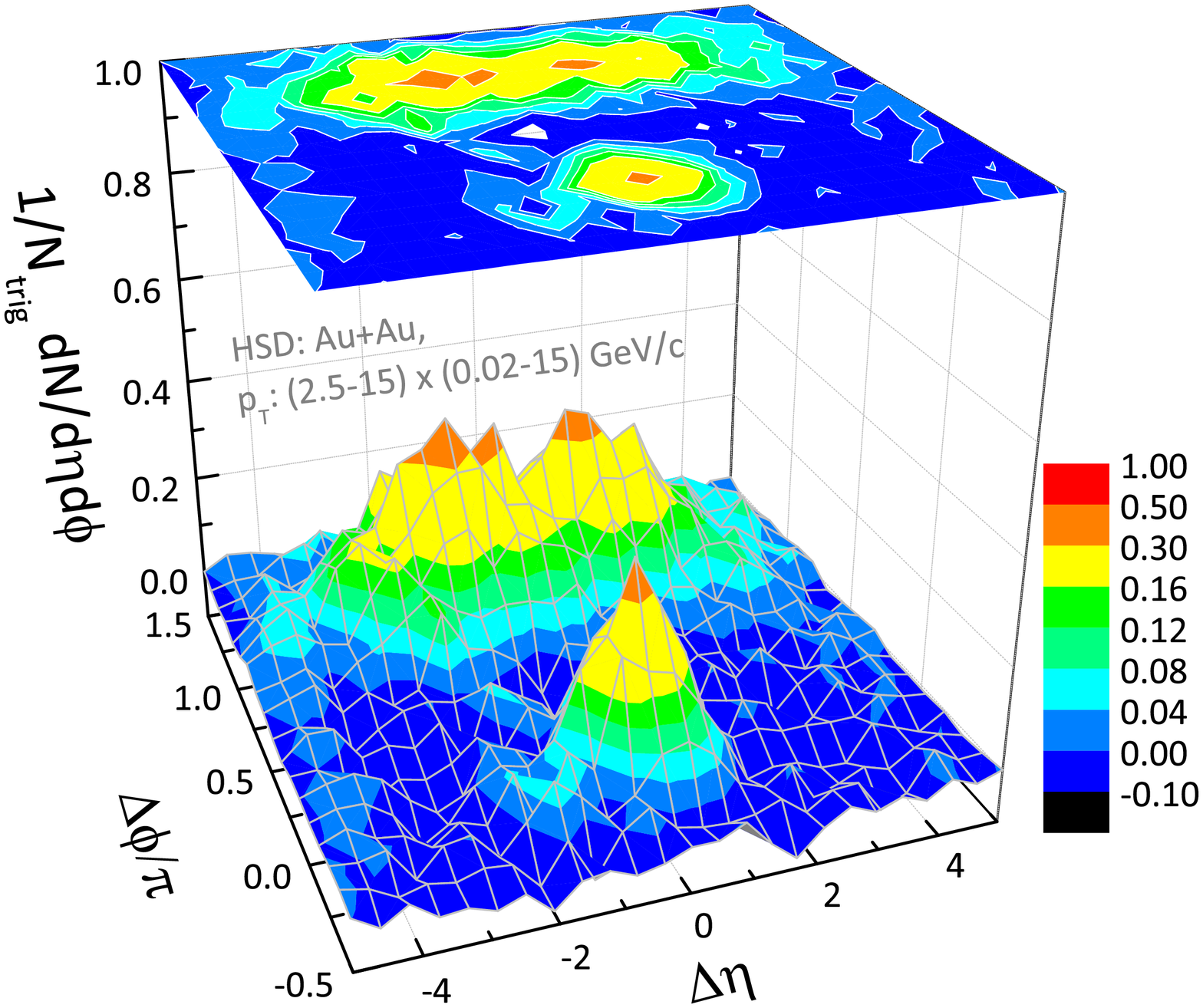,width=0.49\textwidth}
\caption{
The associated particle $(\Delta\eta~,~\Delta\phi)$
distribution (\ref{distr3D}) for p+p ({\it top} panels) and
central Au+Au (b=0, {\it bottom} panels) collisions for the
trigger hadron with $4>p_T^{trig}>6$~GeV/c ({\it left}) and
$p_T^{trig}>2.5$~GeV/c ({\it right}) within the HSD transport approach.}
\label{fig3D}
\end{figure}

%
In Fig.~\ref{fig3D} we present the HSD results for p+p and Au+Au
collisions for the associated differential particle
$(\Delta\eta~,~\Delta\phi)$ distribution (\ref{distr3D}).  We use the
same cuts as the STAR Collaboration, $4<p_T^{trig}<6$~GeV/c and
$2<p_T^{assoc}<4$~GeV/c ~\cite{STAR}, and for the PHOBOS Collaboration,
$p_T^{trig}>2.5~GeV$ and $p_T^{assoc}>0.02$~GeV/c~\cite{PHOBOS}.  In
the HSD transport calculations we obtail on average 0.5 and 5 trigger
particles in an event for the STAR and PHOBOS set of cuts,
correspondingly. The di-jet correlations obtained in the HSD transport
simulations of Au+Au collisions (Fig.~\ref{fig3D}, {\it bottom} panels)
do not show a ridge structure in the pseudorapidity for the near-side jet
as in the data~\cite{STAR,PHOBOS}.

\vspace{0.3cm}
In summary, we conclude that the HSD hadron-string medium does not show
enough suppression for the nuclear modification factor $R_{AA}(p_T)$ at
high $p_T$ and for the away-side jet-associated particles.  For the
first time the medium response on the interactions has been taken into
account in the present non-perturbative HSD calculations in extenstion
to previous perturbative studies \cite{jetHSD}. The non-perturbative
calculations, however, do not reproduce the experimentally observed
`Mach-cone' structure in $\Delta \phi$ for the away-side jet and the
long-range rapidity correlations (the `ridge') for the near-side jet
while supporting the results from perturbative investigations.  It is
interesting to check in future whether the recently proposed parton-HSD
model (PHSD) \cite{pHSD} -- incorporating explicit partonic degrees
of freedom and dynamical hadronization   -- will be able to improve an
agreement with the data and reproduce the structures observed by the
PHOBOS and STAR Collaborations.

\vspace{0.5cm} {\bf Acknowledgements}

We like to thank M.~Ga\'zdzicki and W.~Greiner for useful discussions.
This work was supported by the Helmholtz International Center for FAIR
within the framework of the LOEWE program (Landesoffensive zur
Entwicklung Wissenschaftlich-\"Okonomischer Exzellenz) launched by the
State of Hesse.



\begin{thebibliography}{00}

\bibitem{Baier:1996sk}
  R.~Baier, Y.~L.~Dokshitzer, A.~H.~Mueller, S.~Peigne and D.~Schiff,
  Nucl.\ Phys.\  B {\bf 484}, 265 (1997)
\bibitem{Gyulassy:2003mc}
  M.~Gyulassy, I.~Vitev, X.~N.~Wang and B.~W.~Zhang,
  nucl-th/0302077.
\bibitem{Kovner:2003zj}
  A.~Kovner and U.~A.~Wiedemann,
  hep-ph/0304151.
\bibitem{Adler:2002tq}
  C.~Adler {\it et al.}  [STAR Collaboration],
  Phys.\ Rev.\ Lett.\  {\bf 90}, 082302 (2003).

\bibitem{phenix}
A. Adare et al. [PHENIX collaboration] Phys. Rev. C {\bf 78}, 014901 (2008).


\bibitem{STAR}
  B. I. Abelev et al., [STAR collaboration] Phys. Rev. C  {\bf 80}, 064912 (2009);
  M.~van Leeuwen  [STAR collaboration],
  Eur.\ Phys.\ J.\  C {\bf 61}, 569 (2009).

\bibitem{PHOBOS}
B.~Alver {\it et al.}  [PHOBOS Collaboration],
  Phys.\ Rev.\ Lett.\  {\bf 104}, 062301 (2010).

\bibitem{model_a}
 N. Armesto et al., Phys. Rev. Lett. {\bf 93}, 242301 (2004).

\bibitem{model_b}
 A. Majumder et al., Phys. Rev. Lett. {\bf 99}, 042301 (2007).

\bibitem{model_c}
 C. B. Chiu and R. Hwa, Phys. Rev. C {\bf 72}, 034903 (2005).

\bibitem{model_d}
 C. Y. Wong, Phys. Rev. C  {\bf 78}, 064905 (2008).

\bibitem{model_e}
 S. A. Voloshin, Phys. Lett. B {\bf 632}, 490 (2006);
 E. Shuryak, Phys. Rev. C {\bf 76}, 047901 (2007);
 A. Dumitru et al., Nucl. Phys. A {\bf 810}, 91 (2008);
 K. Dusling et al., Nucl. Phys. A {\bf 828}, 161 (2009);
 J. Takahashi et al., Phys. Rev. Lett. {\bf 103}, 242301 (2009).

\bibitem{Ehehalt}
  W.~Ehehalt and W.~Cassing, Nucl.~Phys.~A {\bf  602}, 449 (1996).
\bibitem{Geiss}
  J.~Geiss, W. Cassing,  and C. Greiner, Nucl.~Phys.~A {\bf 644}, 107 (1998).
\bibitem{Cass99}
  W.~Cassing and E.~L.~Bratkovskaya, Phys.~Rep.~{\bf 308}, 65 (1999).
\bibitem{LUND}
  H.~Pi, Comp.~Phys.~Commun.~{\bf 71}, 173 (1992).
\bibitem{PYTHIA0}
    H.-U.~Bengtsson and T.~Sj\"ostrand,
    Comp.~Phys.~Commun.~{\bf 46}, 43 (1987).

\bibitem{jetHSD}
  W.~Cassing, K.~Gallmeister and C.~Greiner,
  Nucl.\ Phys.\  A {\bf 735}, 277 (2004);
  W.~Cassing, K.~Gallmeister and C.~Greiner,
  J.\ Phys.\ G {\bf 30}, S801 (2004);
  K.~Gallmeister and W.~Cassing,
  Nucl.\ Phys.\  A {\bf 748}, 241 (2005).

\bibitem{pHSD}
  W.~Cassing and E.~L.~Bratkovskaya,
  Phys. Rev. C {\bf 78}, 034919 (2008),
  Nucl.\ Phys.\  A {\bf 831}, 215 (2009).

\end{thebibliography}
\end{document}